\documentclass{WileyMSP-template}
\usepackage[sorting=none]{biblatex} 
\usepackage{amssymb} 
\usepackage{xcolor}
\begin{document}

\pagestyle{fancy}
\rhead{\includegraphics[width=2.5cm]{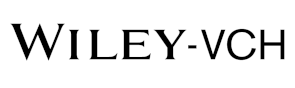}}

\title{Effects of correlated hopping on thermoelectric response of a quantum dot strongly coupled to ferromagnetic leads}

\maketitle


\author{Kacper Wrześniewski}
\author{Ireneusz Weymann}

\begin{affiliations}
Institute of Spintronics and Quantum Information, Faculty of Physics and Astronomy, Adam Mickiewicz University,
Uniwersytetu Poznańskiego 2, 61-614 Poznań, Poland\\
Email Address: wrzesniewski@amu.edu.pl

\end{affiliations}


\keywords{thermoelectric transport; quantum dot; correlated hopping; Kondo effect; exchange field}

\begin{abstract}
We theoretically investigate the impact of correlated hopping on thermoelectric transport through a quantum dot coupled to ferromagnetic leads. Using the accurate numerical renormalization group method, we analyze the transport characteristics, focusing on the interplay between electronic correlations, spin-dependent transport processes, and thermoelectric response. We calculate the electrical conductance and thermopower as functions of the dot energy level, lead polarization, and the amplitude of correlated hopping. Moreover, we analyze the effect of competing correlations on the Kondo resonance and discuss the asymmetry of conductance peaks under the influence of the exchange field.
We demonstrate that the presence of correlated hopping is responsible for asymmetric spin-dependent transport characteristics. Our results provide valuable insight into how correlated hopping affects spin-dependent transport and thermoelectric efficiency
in quantum dot systems with ferromagnetic contacts.
\end{abstract}

\section{Introduction}

Thermoelectric effects play a crucial role in energy conversion and transport in nanoscale and hybrid systems. These effects, arising from the interplay between heat and charge flow, enable direct conversion of temperature gradients into electrical power and vice versa \cite{barnard1972thermoelectricity, Mahan1997Mar, Horodecki2013Jun, YungerHalpern2016Feb, Gelbwaser-Klimovsky2018Apr}.
Thermoelectric devices have an important place in contemporary technology.
They are used for waste heat recovery, as heat engines and to improve energy efficiency.
In addition, thermoelectric coolers are employed in electronic devices,
medical applications, and precision temperature control systems,
demonstrating their versatility in engineering  \cite{Linden2010Sep, Rossnagel2016Apr, Mitchison2019Apr, Bu2022Jan, Smriti2025Jan, Aamir2025Feb}.

Understanding the thermoelectric properties in the context of quantum systems,
such as quantum dots and other nanostructures, is essential to design
efficient energy-harvesting devices and exploring fundamental
aspects of non-equilibrium physics \cite{Segal2005Oct, Giazotto2006Mar, Costi2010Jun, Wysokinski2012Jul, Trocha2012Feb, Chirla2014Jan, Wojcik2014Apr, Sánchez_2014, Benenti2017Jun, Dutta2019Jan, Gorski2019Jul, Eckern2021Oct}.
Theoretical investigations of thermoelectric effects in nanostructures
are of great importance, since these systems are predicted to
exhibit significantly higher thermoelectric efficiency
as compared to bulk materials due to enhanced quantum confinement,
energy filtering, and reduced phonon transport.
Moreover, analyzing thermoelectric coefficients,
such as thermopower, can provide further valuable insights
into the underlying electron correlations,
quantum interference phenomena, and even signatures of many-body effects,
including the Kondo \cite{Kondo, Goldhaber-Gordon1998, KondoQD} or Fano resonances \cite{Fano1961Dec, Sasaki2009Dec, Zitko2010Mar}.

In this work, we study the thermoelectric properties
of a quantum dot strongly coupled to external ferromagnetic leads,
incorporating the effects of correlated hopping \cite{Hubbard1963Nov, Dolcini2013Sep, Foglio1979Dec}.
We note that the analysis of thermoelectric phenomena
in nanoscopic systems becomes especially compelling
when magnetic subsystems are involved \cite{Koch04, Krawiec2006Feb, Swirkowicz2009Nov, Wang10, Misiorny12, Rejec2012Feb, Weymann2013Aug, Misiorny14, Wojcik15, Manaparambil2021Apr, Liu2023Sep, Trocha2025Feb}.
The discovery of the spin Seebeck effect has led to the rapid growth
of the field of spin caloritronics, which explores the intricate interplay between
charge, heat, and spin transport. In such systems,
thermoelectric coefficients become spin-dependent,
leading to novel and rich behavior not seen in non-magnetic counterparts.
Moreover, in correlated magnetic nanostructures,
such as quantum dots coupled to ferromagnetic leads, the spin thermopower
has proven to be a sensitive probe of exchange fields
and their interaction with electronic correlations,
including those responsible for the Kondo effect \cite{Weymann2013Aug}.
Additionally, the spin-dependent thermoelectric effects have
been investigated in quantum dots under external magnetic fields \cite{Costi19, Crisan2020Oct, BibEntry2025Jan},
revealing further complexity and potential for tunability.
This work aims to deepen our understanding of thermoelectric
behavior in strongly correlated quantum dots by focusing
on spin-resolved transport in the presence of correlated hopping.
To achieve this goal, we employ the nonpreturbative
numerical renormalization group (NRG) method \cite{Wilson1975Oct, Bulla2008Apr},
which allows for exploration of various correlations
in a very accurate manner.

The paper is structured as follows. Section \ref{sec:tf} presents the theoretical framework,
in which we describe the model and briefly introduce the NRG method. In the main section \ref{sec:results} we present and discuss  in detail the obtained results.
Finally, we conclude the paper in the last section \ref{sec:conclusion}.

\section{Theoretical framework}\label{sec:tf}
\subsection{Hamiltonian}

\begin{figure}
\centering
  \includegraphics[width=0.65\linewidth]{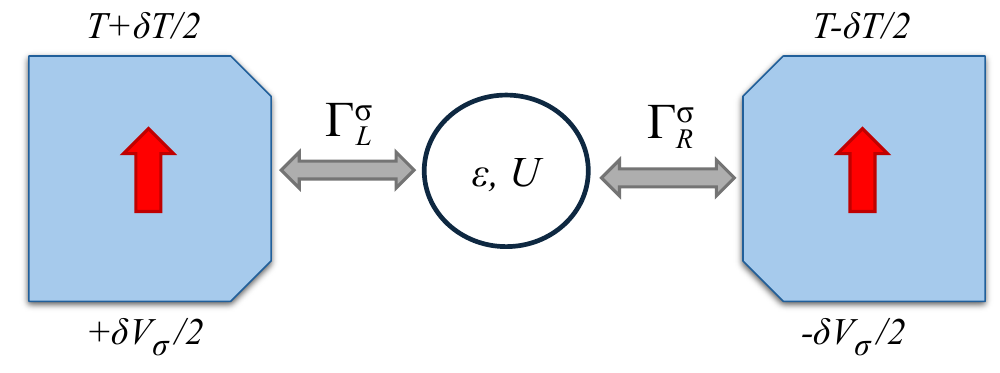}
  \caption{Schematic of the considered quantum dot system with Coulomb interaction $U$ and dot level energy $\varepsilon$.
  The central part is connected to two ferromagnetic leads with coupling strength $\Gamma^\sigma_{L/R}$, for the left and right lead. 
  There is a temperature gradient $\delta T$ and a small
  voltage gradient $\delta V_\sigma$ applied symmetrically to the system.}
  \label{fig:fig1}
\end{figure}

We consider a single level quantum dot (QD) strongly coupled to two ferromagnetic leads
with parallel alignment of magnetizations, as depicted in Fig. \ref{fig:fig1}.
This system can be described by the single impurity Anderson
Hamiltonian of the following form
\begin{equation}
    H =  H_{Leads} + H_{QD} + H_T,
    \label{eq:H_total}
\end{equation}
where the ferromagnetic leads are modeled by first term
\begin{equation}
    H_{Leads} = \sum_{\alpha \sigma \bf{k}} \varepsilon_{\alpha \sigma \bf{k}} c^{\dagger}_{\alpha \sigma \bf{k}} c_{\alpha \sigma \bf{k}},
\label{eq:H_leads}
\end{equation}
and the operator $c^{\dagger}_{\alpha \sigma \bf{k}} (c_{\alpha \sigma \bf{k}})$ creates (destroys) an electron with spin $\sigma$ and energy $\varepsilon_{\alpha \sigma \bf{k}}$
in lead $\alpha = L/R$ (left/right).
The isolated single-level QD is described by
\begin{equation}
    H_{QD} =  \sum_\sigma \varepsilon d^{\dagger}_\sigma  d_\sigma + U d^{\dagger}_\uparrow  d_\uparrow d^{\dagger}_\downarrow  d_\downarrow,
\label{eq:H_qd}
\end{equation}
and Hamiltonian
\begin{equation}
    H_{T} =  \sum_{\alpha \sigma \bf{k}} V_{\alpha \sigma \bf{k}}(1 - x d^\dagger_{\bar{\sigma}} d_{\bar{\sigma}})c^{\dagger}_{\alpha \sigma \bf{k}}d_\sigma + H. c.,
\label{eq:H_qd}
\end{equation}
specifies tunneling including correlated hopping. Here, $d^{\dagger}_\sigma$ ($d_\sigma$) creates (destroys) electron with spin $\sigma$ and energy $\varepsilon$ on QD,
$U$ denotes the on-site Coulomb interaction, $V_{\alpha \sigma \bf{k}}$ is the tunnel matrix element between QD and $\alpha$-lead, 
assumed to be momentum-independent $V_{\alpha \sigma \bf{k}} \equiv V_{\alpha \sigma}$, 
and $x$ is the correlated hopping parameter.
The spin-dependent coupling between QD and FM lead is given by $\Gamma^\sigma_\alpha = (1\pm p_\alpha)\Gamma_\alpha$, where $p_\alpha$ is the spin polarization of lead $\alpha$,
defined as $p_\alpha = (\Gamma^\uparrow_\alpha - \Gamma^\downarrow_\alpha)/(\Gamma^\uparrow_\alpha + \Gamma^\downarrow_\alpha)$, with $\Gamma^\sigma_\alpha=\pi \rho^\sigma_\alpha V^2_{\alpha \sigma}$, $\rho^\sigma_\alpha$ being the spin-dependent density of states of respective lead and $\Gamma_\alpha = (\Gamma^\uparrow_\alpha + \Gamma^\downarrow_\alpha)/2$.
Moreover, for the NRG calculations, we perform the left-right orthogonal transformation, after which the quantum dot couples to an effective conduction channel with coupling strength $\Gamma \equiv \Gamma_L + \Gamma_R$, with an effective spin polarization $p\equiv (p_L+p_R)/2$.

\subsection{Transport coefficients}

We focus on the linear response regime with respect to the applied bias and temperature gradients.
Then, the linear transport coefficients can be related to the following Onsager integrals
\begin{equation}
    L_{n \sigma} = - \frac{1}{h}\int d \omega \: \omega^n \frac{\partial f(\omega)}{\partial \omega} T_\sigma (\omega),
\label{eq:L_general}
\end{equation}
with $T_\sigma(\omega)$ being the transmission function and $f(\omega)$ standing for the Fermi-Dirac distribution. The transmission function is obtained from the corresponding retarded Green's function of the quantum dot, which is calculated in the Lehmann representation directly from the discrete NRG spectrum \cite{Bulla2008Apr}.
Using the above coefficients, one can find the spin-resolved linear conductance
\begin{equation}
    G_\sigma = e^2L_{0\sigma},
\label{eq:G_general}
\end{equation}
and the corresponding total conductance given by $G=G_\uparrow+G_\downarrow$.
The Seebeck coefficient assuming absence of spin accumulation in the contacts is given by \cite{Swirkowicz2009Nov,Weymann2013Aug}
\begin{equation}
    S \equiv - \left( \frac{\delta V}{\delta T} \right) _{J=0} =-\frac{1}{|e|T}\frac{L_1}{L_0},
\label{eq:S_general}
\end{equation}
where $L_{n}=L_{n \uparrow}+L_{n \downarrow}$. 
Finally, when the electrodes have long spin relaxation time,
such that the spin accumulation may develop,
the spin Seebeck effect can emerge. The spin thermopower can be evaluated from
\begin{equation}
    S_S \equiv - \left( \frac{\delta V_\sigma}{\delta T} \right) _{J_\sigma=0} =-\frac{2}{\hbar T}\frac{M_1}{L_0},
\label{eq:Ss_general}
\end{equation}
with $M_1 = L_{1 \uparrow} - L_{1 \downarrow}$. In the above, $\delta V$ ($\delta V_\sigma$) is a small (spin) voltage gradient applied symmetrically to the system, as indicated in Fig. \ref{fig:fig1}.

\subsection{Method}

To obtain reliable and experimentally testable predictions for the transport properties of a QD coupled to ferromagnetic leads, including the effect of correlated hopping, we employ the full density matrix NRG method (fDM-NRG) \cite{Wilson1975Oct, Bulla2008Apr} implemented in the open access Budapest Flexible NRG code \cite{Legeza2008Sep}.
This approach allows us to accurately analyze the local density of states
and thermoelectric transport properties of the examined system
across the full range of model parameters.

In the NRG framework, the conduction band is logarithmically discretized,
and the system's Hamiltonian is transformed into a tight-binding chain with exponentially decaying hoppings. This chain is then diagonalized iteratively. In our calculations, we retained at least $2000$ states per iteration to ensure numerical accuracy.

As mentioned above, to facilitate the analysis, we apply the orthogonal left-right transformation, mapping the original two-lead Hamiltonian to an effective form in which the quantum dot couples exclusively to an even-parity combination of electron operators from the left and right leads. The resulting coupling strength is then given by $\Gamma=\Gamma_L+\Gamma_R$. 
Generally, the magnetic moments of the leads can form two magnetic configurations, the parallel one, when the moments point in the same direction
and the antiparallel one, when the magnetic moments point in the opposite direction. However, since the most interesting spin-resolved effects 
are present in the parallel configuration of the system \cite{Gaass2011Oct},
in this work we focus on this magnetic alignment, cf. Fig.~\ref{fig:fig1}.
We also note that the performed orthogonal transformation
for the considered system, and for the assumed parallel alignment of the leads' magnetic moments, is not restricted to symmetric systems,
but is also valid in the case of asymmetry, 
with the effective coupling parameters being $\Gamma$ and $p$.

\section{Numerical results and discussion}\label{sec:results}

In this section, we present and discuss the main results
of the paper. In particular, we study the temperature
and dot level dependence of the linear conductance and the Seebeck coefficient
exploring the effects of correlated hopping.
We also examine the behavior of the spin Seebeck coefficient.

\subsection{Temperature dependence of linear conductance and thermopower}

When analyzing the temperature dependence of transport properties in quantum dot systems, it is essential to consider the relevant energy scales. In the singly occupied regime ($0>\varepsilon/U>-1$), corresponding to the odd Coulomb valley, the dominant energy scale is associated with the Kondo temperature $T_K$, which governs the emergence of Kondo correlations. The Kondo temperature in the case of $x=0$, which will be used as a reference, can be evaluated from
\begin{equation}
    T_K\approx\sqrt{\frac{U\Gamma}{2}} \exp{\left[\frac{\pi\varepsilon(\varepsilon+U)}{2\Gamma U}\frac{\rm{arctanh}(p)}{p}\right]},
\end{equation}
as obtained using the Haldane scaling theory \cite{Anderson1970Dec,Haldane1978Feb,Martinek2003Sep}. 
Additionally, when the quantum dot is connected to ferromagnetic electrodes, another energy scale appears, associated with
the so-called effective exchange field, which 
at low temperatures and for $x=0$ is approximately equal to \cite{Martinek2003Sep}
\begin{equation}
    \Delta \varepsilon_{exch}=\frac{2p\Gamma}{\pi}\ln \left|\frac{\varepsilon}{\varepsilon+U} \right|.
\end{equation}
This effect is arising from the spin-dependent tunneling,
which can significantly influence the spin dynamics
and low-temperature behavior of the system \cite{Gaass2011Oct,Martinek2003Sep,Martinek2003Dec}.
It is important to note here that, as follows from the above formula,
the exchange field vanishes 
at the particle-hole symmetry point, i.e. $\varepsilon=-U/2$,
where the Kondo resonance remains unaffected by leads' ferromagnetism.
When the system is detuned from this point, the Kondo peak becomes split.
However, this becomes modified in the presence of assisted hopping,
which generally breaks the particle-hole symmetry of the system,
as demonstrated in the sequel. As a final remark, we point out that introducing spin-dependent tunneling due to ferromagnetic electrodes breaks the known symmetry in quantum dots with correlated hopping, where the transformation $x \rightarrow 2-x$ preserves occupation number $\langle n \rangle$ and transport properties \cite{Tooski_2014}. We checked numerically that, although the occupation number is only slightly affected over a broad range of $x$, the transport coefficients are quantitatively altered by this transformation.

\begin{figure}[h]
\centering
  \includegraphics[width=0.6\linewidth]{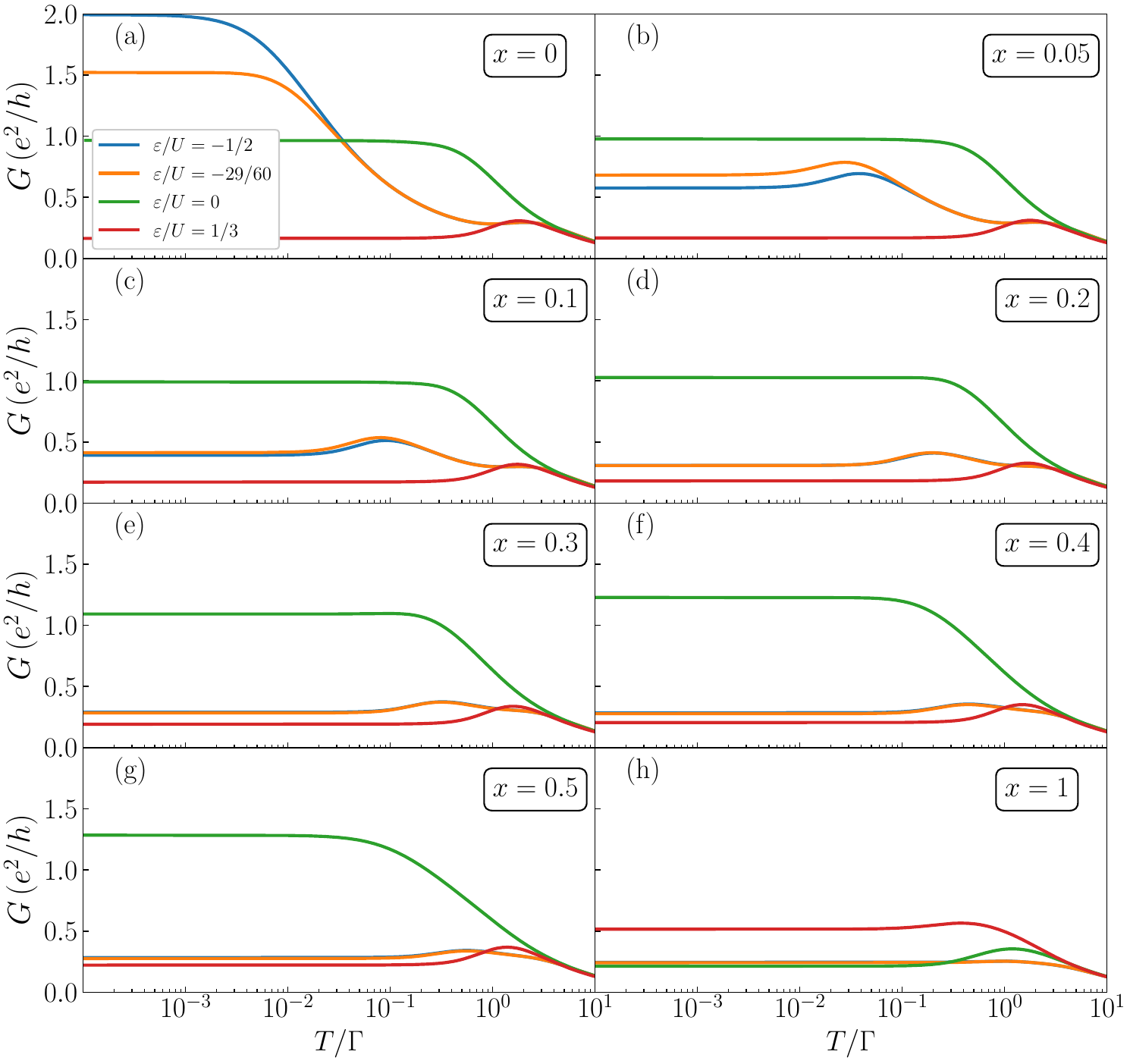}
  \caption{The temperature dependence of the linear conductance $G$ 
for different values of the level position $\varepsilon$, as indicated in the legend. (a)-(h) present the results for different values of the correlated hopping parameter $x$, as indicated in the figure. The parameters are: $U=0.1D$, $\Gamma/U=0.1$ and $p=0.5$, where $D$ denotes the band halfwidth, which is used as energy unit.}
  \label{fig:fig2}
\end{figure}

We start the discussion of numerical results with the analysis of the linear conductance $G$ as a function of temperature, which is shown in Fig.~\ref{fig:fig2}. In the following, for the ferromagnetic leads we assume the spin polarization $p=0.5$, unless stated otherwise. When correlated hopping is not present ($x=0$), typical dependence is observed in the Kondo regime, with an increase of the conductance to the maximum ($G=2e^2/h$)
for the orbital level tuned to the particle-hole symmetry point ($\varepsilon/U=-0.5$). As the energy of the QD level is shifted away from this point,
the Kondo effect gets suppressed due to the presence of the exchange field \cite{Martinek2003Sep, Pasupathy2004Oct,Weymann2011}.

When the correlated hopping is switched on ($x>0$), one can see an immediate suppression
of the Kondo resonance at the particle-hole symmetry point,
as the conductance drops strikingly in this regime
even for relatively small values of $x$. 
This is caused by the fact that assisted hopping breaks the
particle-hole symmetry and thus the exchange field may also occur for $\varepsilon/U=-0.5$ suppressing the conductance through the system.
On the other hand, in the mixed-valence regime ($\varepsilon=0$),
the constructive effect of correlated hopping is revealed, as the linear conductance dependence is undergoing considerable amplification. 
Furthermore, the linear conductance dependence is not affected
for the system with almost empty orbital level
($\varepsilon/U=1/3$), until $x=1$ is reached. 
For this specific value of correlated hopping,
the decoupling of doubly occupied states occurs, resulting in a drop of the linear conductance for a wide range of orbital level detuning. 
The intriguing increase of linear conductance for $\varepsilon/U=1/3$ is due to the asymmetric influence of correlated hopping on the Kondo effect,
which is dimmed and shifted toward the upper Hubbard band.

\begin{figure}[t]
\centering
  \includegraphics[width=1\linewidth]{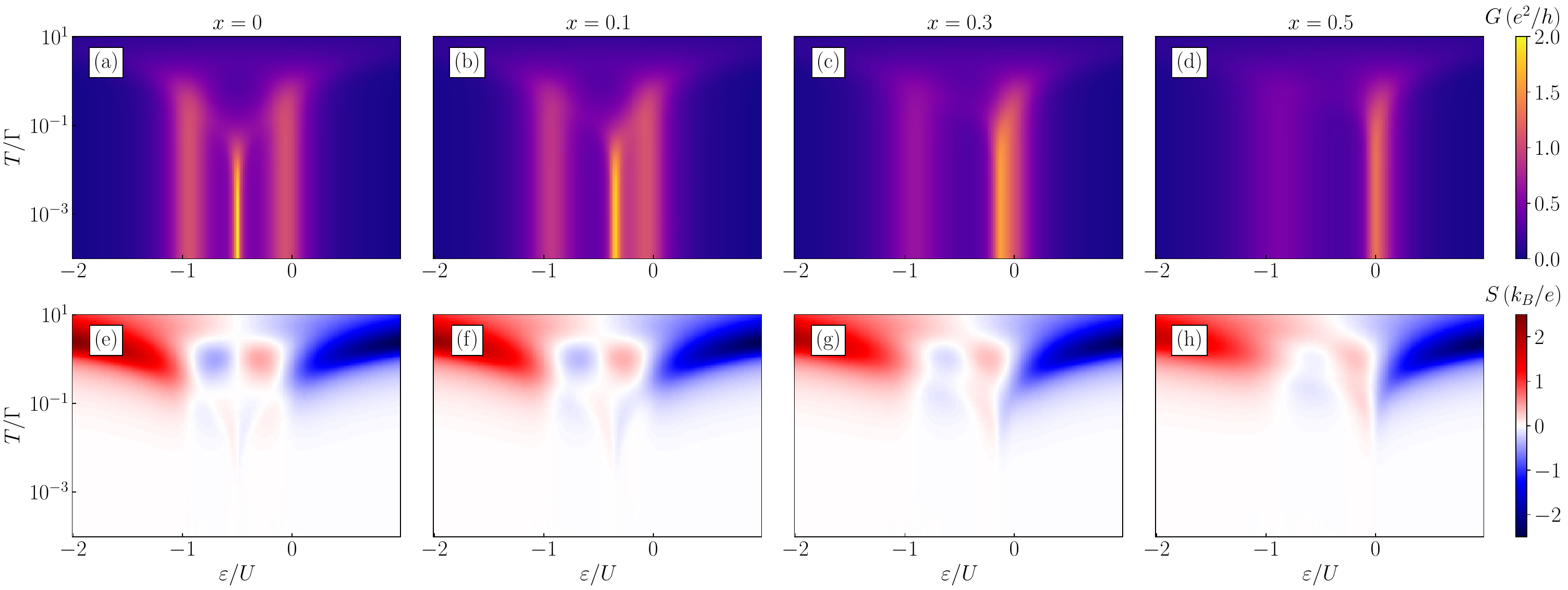}
  \caption{The temperature and energy level dependence of the linear conductance $G$ (upper row) and thermopower $S$ (lower row).
  The first column (a,e) shows the results in the absence of correlated hopping, while (b,f) present results for $x=0.1$,
  (c,g) show results for $x=0.3$,
  and (d,h) display results for $x=0.5$.
  The other parameters are the same as in Fig.~\ref{fig:fig2}.}
  \label{fig:fig3}
\end{figure}

The evolution of this effect is systematically presented in Fig.~\ref{fig:fig3}. 
The first row presents the linear conductance $G$, 
while the second row shows the thermopower $S$
as functions of temperature $T$ and energy level $\varepsilon$.
Let us first consider the case when the correlated hopping is not present
[$x=0$, see panel (a) and (e)]. As can be seen, in the low-temperature regime, the Kondo resonance develops together with two Coulomb peaks
symmetrically located on
both sides with respect to the particle-hole symmetry point $\varepsilon/U=-0.5$.
Once the correlated hopping is switched on and increased, the maximum conductance
is suppressed and moves toward the upper Hubbard resonance, achieving mixed-valence regime for $x=0.5$.
A similar effect was previously predicted and explained as a consequence of particle-hole symmetry breaking induced by correlated hopping in the case of nonmagnetic systems \cite{Gorski2019Jul,Eckern2021Oct,Tooski_2014}.
We recall that, in the absence of correlated hopping, the system coupled to ferromagnetic leads reveals splitting of the Kondo resonance
when quantum dot is detuned away from the particle-hole symmetry point.
Here, we show that when the symmetry is broken due to the presence of correlated hopping,
the Kondo resonance survives and is pushed away from $\varepsilon/U=-0.5$.

The associated thermopower $S$ reveals oscillatory behavior and multiple sign changes with the positive values indicating transport dominated by holes, while the negative values---by electrons. When correlated hopping is absent [see panel (e)], in the high temperature regime ($T/\Gamma \gtrsim 10^{-1}$),
there are three occurrences of sign change as the energy of orbital level is swept. When the temperature is lowered and the Kondo correlations are becoming significant, the overall thermopower is considerably suppressed,
but two more sign changes are present.
Upon introduction of correlated hopping, the oscillatory behavior
as a function of $\varepsilon$ with multiple sign changes 
is dominantly present well above the Kondo temperature, 
with weak quantitative differences.
However, for lower temperatures, the sign change associated
with position of the Kondo resonance is shifted accordingly,
and the symmetry of the overall dependence gets strongly distorted.

\subsection{Level detuning and spin polarization dependence}

\begin{figure}[h]
\centering
  \includegraphics[width=0.6\linewidth]{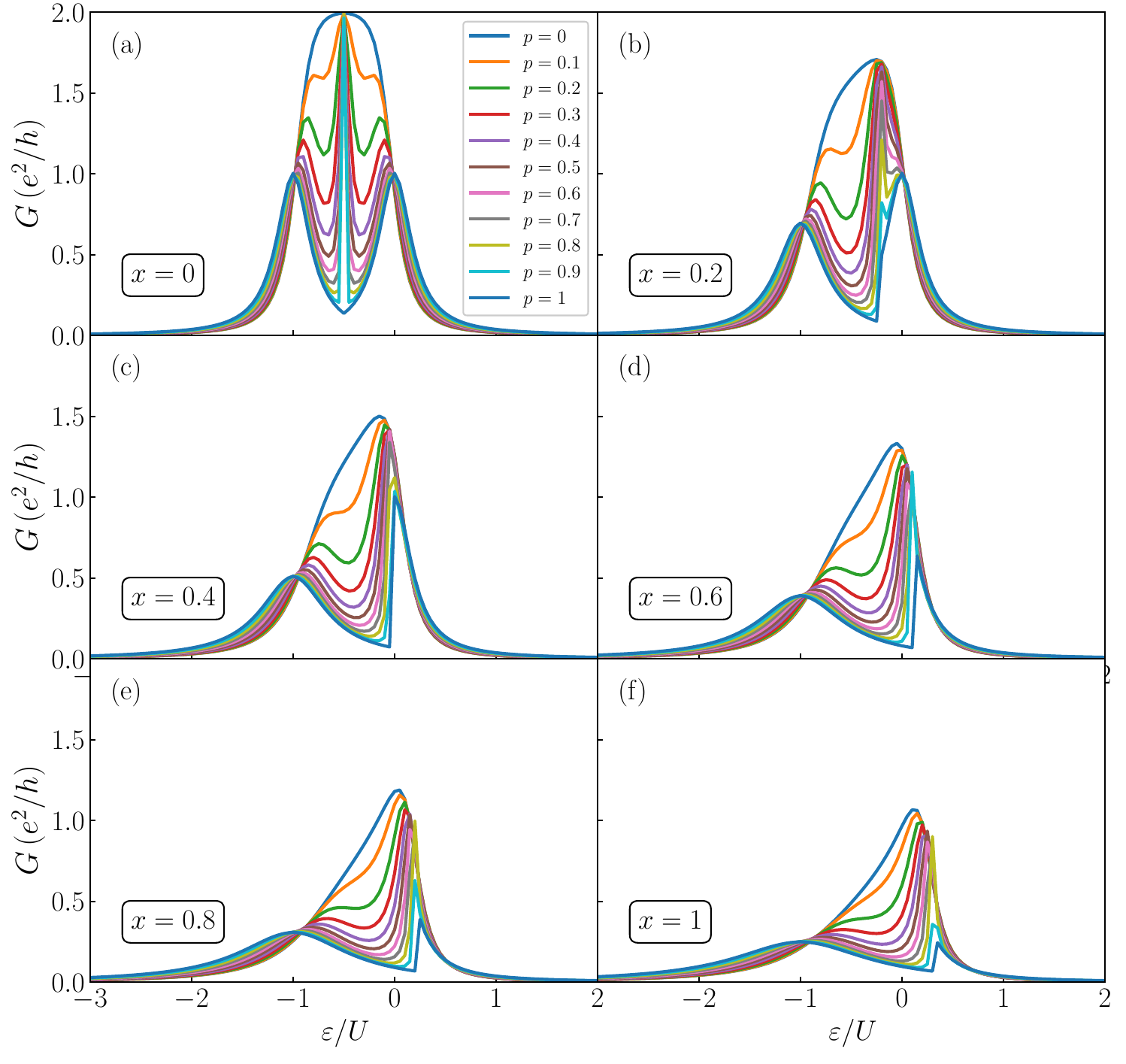}
  \caption{The linear conductance $G$ as a function of the level position $\varepsilon/U$ for different values of
  the spin polarization $p$, as indicated.
  (a)-(f) present the results for selected values of 
  the correlated hopping parameter $x$.
  In calculations we assumed $T/U=10^{-5}$, while other parameters are
  the same as in Fig.~\ref{fig:fig2}.}
  \label{fig:fig4}
\end{figure}

In Fig. \ref{fig:fig4} we present the dependence of the linear conductance as a function of orbital level energy $\varepsilon$, for different values of lead spin polarization and correlated hopping parameter in the low temperature regime $T/U=10^{-5}$, i.e. for $T \lesssim T_K$.
For $x=0$ and $p=0$, see Fig.~\ref{fig:fig4}(a), the Kondo effect is fully developed. However, as spin polarization is raised to non-zero values, a typical behavior of impurity coupled to ferromagnetic lead is observed with suppression and splitting of the Kondo peak when the system is detuned from the particle-hole symmetry point. When the correlated hopping is present in the system,
the level detuning dependence is strongly modified.
The maximum of the linear conductance is lowered as $x$ acquires higher values,
and the Kondo peak is shifted toward the mixed-valence regime ($\varepsilon=0$). 
Additionally, when the ferromagnetic electrodes have a stronger spin polarization, 
the linear conductance is even further suppressed,
specifically in the whole range of the Coulomb valley.
We note that for the assumed low temperature,
in accordance with the Sommerfeld expansion, $S\sim T$,
the thermopower $S$ is close to zero
in the whole range of level detuning.
Therefore, we will show its dependence only for higher temperatures.

\begin{figure}
\centering
  \includegraphics[width=0.6\linewidth]{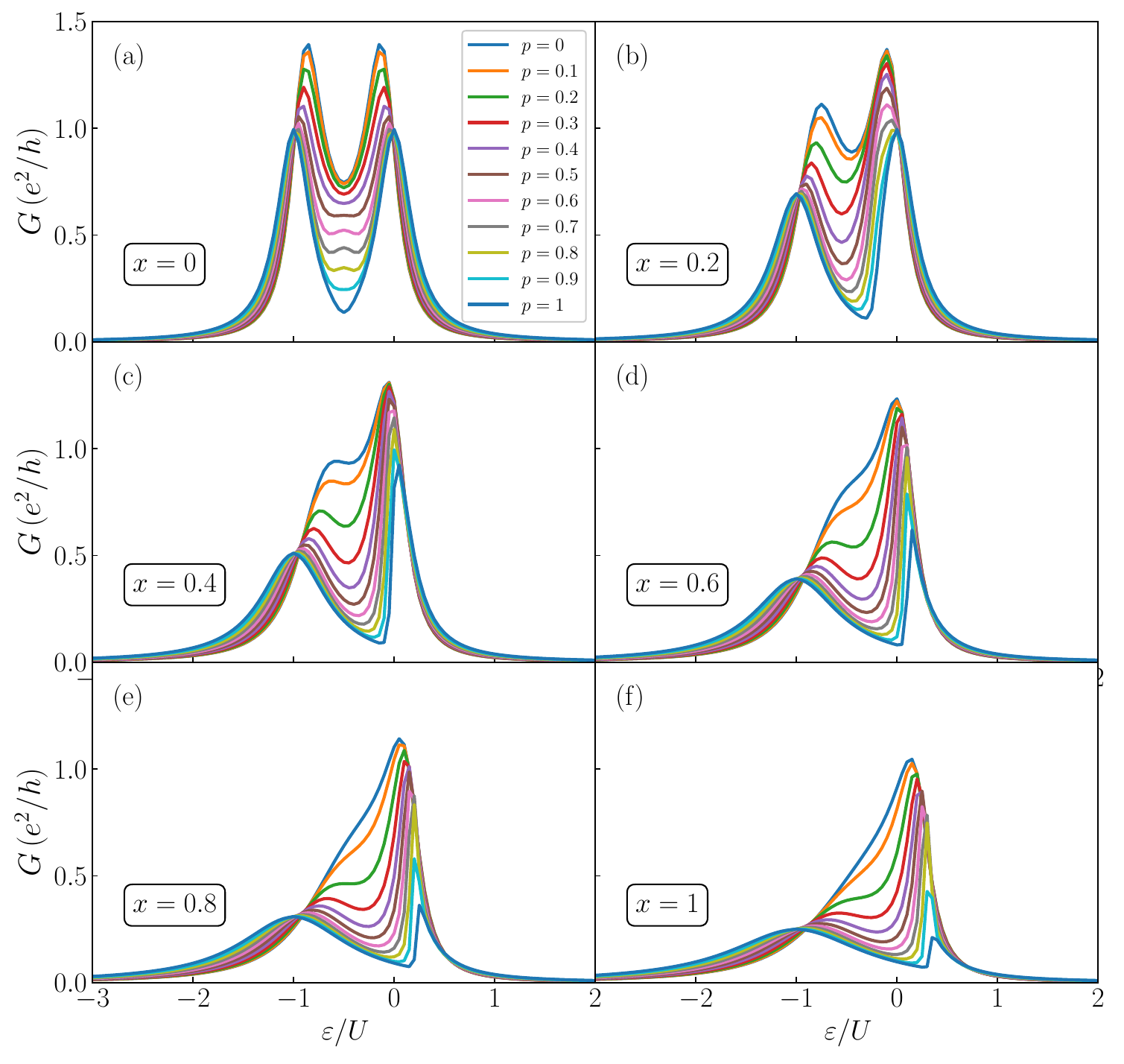}
  \caption{
  The linear conductance $G$ as a function of the level position $\varepsilon/U$ for different values of
  the spin polarization $p$, as indicated.
  (a)-(f) present the results for selected values of 
  the correlated hopping parameter $x$.  In calculations we assumed $T/U=10^{-2}$, while other parameters are
  the same as in Fig.~\ref{fig:fig2}.}
  \label{fig:fig5}
\end{figure}

Subsequently, in Fig. \ref{fig:fig5} we present the linear conductance as a function of the orbital level energy and spin polarization for six different values of the correlated hopping parameter. The analysis is performed in a similar manner to that shown previously in Fig.~\ref{fig:fig4}, but here we focus on a higher temperature of $T/U=10^{-2}$, where thermoelectric effects become significant. Several important observations can be made. In panel (a), where no correlated hopping is present, the characteristic Kondo effect and associated peak at the particle-hole symmetric point are absent; instead, only two symmetric peaks corresponding to the Coulomb blockade edges are visible. This is obviously due to the fact that the assumed temperature is higher than the Kondo temperature.
As the spin polarization increases, both peaks exhibit a notable reduction in their maximum values, and the minimum of conductance at $\varepsilon/U=-0.5$ becomes deeper. Upon introducing correlated hopping, the system's behavior changes significantly: the lower-energy peak becomes strongly suppressed independently of spin polarization $p$.
Such asymmetric height of the lower and upper Hubbard peaks was
already considered by Eckern and Wysokiński \cite{Eckern2021Oct}
as a possible fingerprint of correlated hopping present in the system,
which could be verified experimentally.
Our results provide even stronger indication of the presence of the considerable exchange field for values up to $x=0.5$.
When the spin polarization of the leads is increased,
so is the difference between the lower and upper Hubbard peaks.
Additionally, by decreasing the minimum in the dependency, the exchange field
can facilitate distinction of the modified resonances in the regime,
where strong correlated hopping smears out the two-peak structure.
Finally, we note that high values of spin polarization
can influence the overall conductance by strongly
suppressing one of the spin channels and, therefore,
optimal values for such purpose should be between $0.3 \lesssim p \lesssim 0.5$.

\begin{figure}
\centering
  \includegraphics[width=0.6\linewidth]{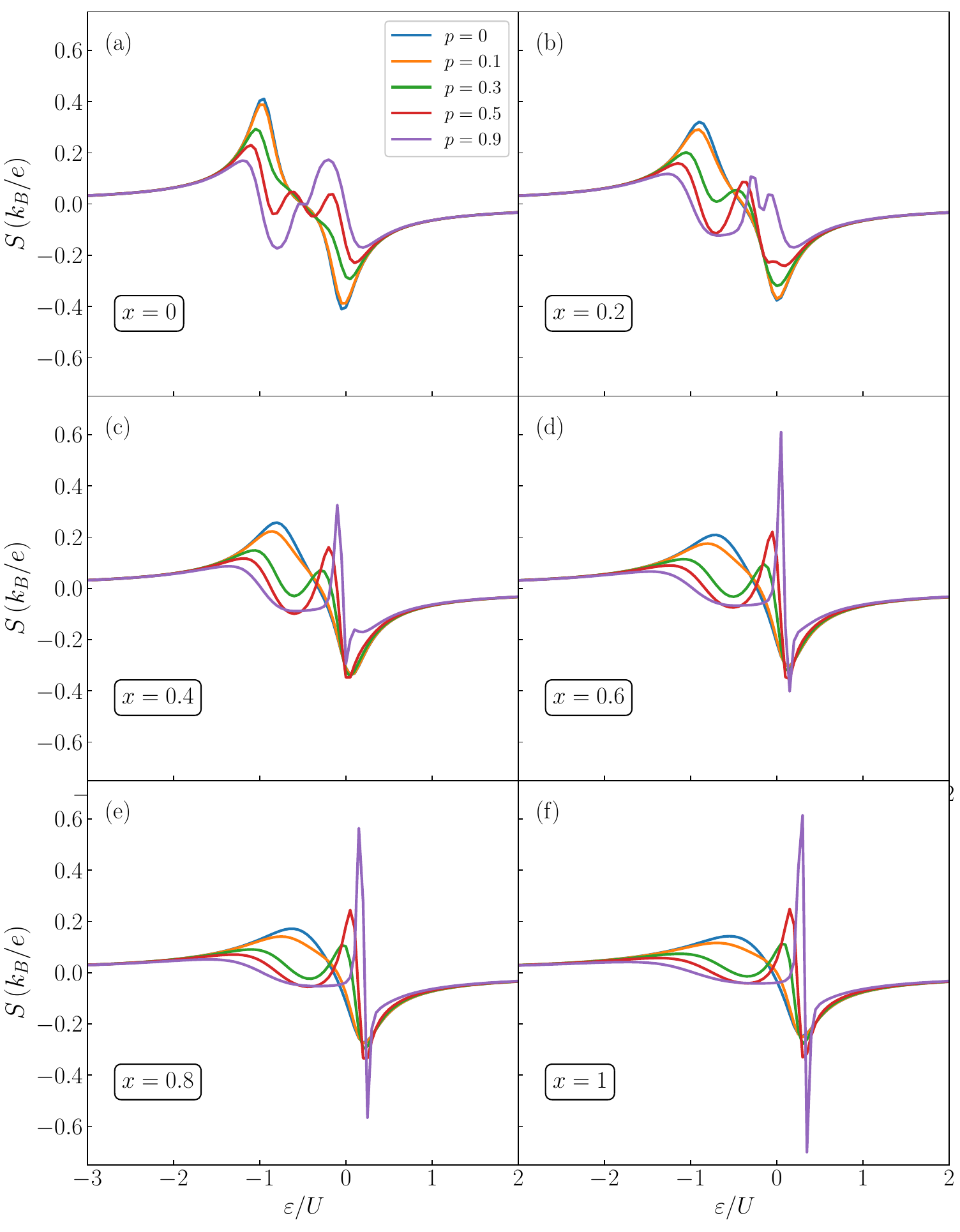}
  \caption{
  The dependence of the thermopower $S$ on the quantum dot
  level position $\varepsilon/U$ for different values of
  the spin polarization $p$, as indicated.
  (a)-(f) present the results for selected values of 
  the correlated hopping parameter $x$.
  In calculations we assumed ${T/U=10^{-2}}$,
  while the other parameters are as in Fig.~\ref{fig:fig2}.}
  \label{fig:fig6}
\end{figure}

Furthermore, we present the thermopower $S$ as a function of the orbital level energy, with different panels, see Figs.~\ref{fig:fig6} (a)–(f),
corresponding to increasing values of the correlated hopping parameter.
In the absence of spin polarization and correlated hopping, 
the thermopower displays a characteristic behavior:
a pronounced maximum near the lower Hubbard peak,
a sign change at the particle-hole symmetry point,
and a symmetric minimum near the right Hubbard resonance.
As the spin polarization increases,
the absolute values of both extrema decrease,
and for $p \gtrsim 0.5$, an additional sign change appears
in the half-filling regime, reflecting a more complex particle-hole asymmetry. Moreover, the presence of correlated hopping significantly modifies the behavior of the thermopower, primarily for $\varepsilon/U>-0.5$.
In particular, for strong spin polarization values, a Fano-like
profile emerges around the orbital energy of $\varepsilon/U=1/3$,
accompanied by a marked enhancement of the thermopower's absolute value,
which can exceed $S = k_B/2e$.

\begin{figure}
\centering
  \includegraphics[width=0.6\linewidth]{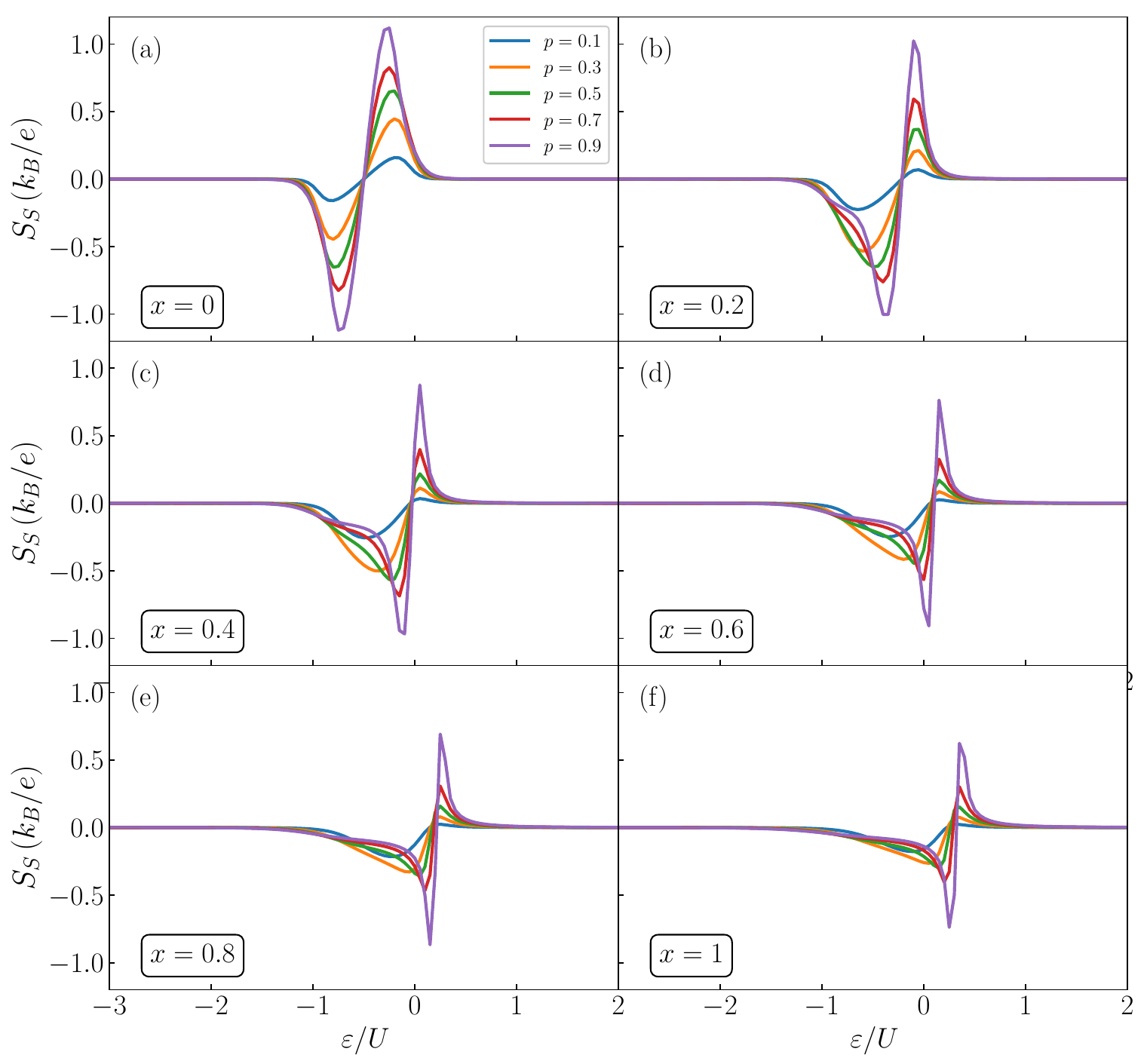}
  \caption{
  The spin thermopower $S_S$ as a function of the quantum dot
  level position $\varepsilon/U$ calculated for different values 
  of the spin polarization $p$, as indicated.
  (a)-(f) display the results for selected values of 
  the correlated hopping parameter $x$.
  In calculations we assumed $T/U=10^{-2}$ and the
  other parameters are the same as in Fig.~\ref{fig:fig2}.}
  \label{fig:fig7}
\end{figure}

To provide a more detailed view of the influence of spin polarization in the ferromagnetic electrodes and spin-dependent effects in the system, we now analyze the spin thermopower $S_S$. Such thermopower can emerge in the system when there is a spin accumulation in the leads that gives rise to a spin bias \cite{Swirkowicz2009Nov}. 
The corresponding results are presented in Fig.~\ref{fig:fig7}.
Across a wide range of parameters, the spin thermopower shows
a characteristic dip-peak structure as a function of energy level. 
Importantly, with increasing the spin polarization $p$,
both dip and peak become more pronounced, leading to higher absolute
values of the extrema. On the other hand, when correlated hopping
is introduced and subsequently increased, the overall magnitude of $S_S$ decreases, reflecting the suppression of spin-dependent thermoelectric effects.
In addition, the positions of the dip, peak, and the associated sign change shift systematically towards higher values of the orbital energies.
For sufficiently large correlated hopping and strong spin polarization,
the dip-peak structure becomes tightly compressed and centered
around an orbital energy of approximately $\varepsilon/U=1/3$,
indicating a strong interplay between spin correlations
and interference effects in the system.

\section{Conclusions}\label{sec:conclusion}

We have theoretically analyzed the spin-resolved electric and thermoelectric transport properties of a quantum dot system,
where the dot is coupled to ferromagnetic leads,
in the presence of correlated hopping.
To capture all the many-body correlations accurately, we have employed the numerical renormalization group method, which allowed us for an exact treatment of the system's strongly correlated nature. For quantum dots with ferromagnetic contacts, the Kondo effect 
becomes generally suppressed due to the exchange field, except for the particle-hole symmetry point. Our analysis of the linear conductance revealed that,
in the presence of correlated hopping, the Kondo resonance is shifted toward higher energies of the quantum dot orbital level.
We have also demonstrated the suppression of resonant peaks with increasing the spin polarization, and a strong overall conductance reduction with the introduction of correlated hopping. The asymmetry of modified conductance
peaks is further enhanced by the presence of the exchange field
induced by the coupling of the quantum dot to ferromagnetic leads.
On the other hand, the study of the thermopower demonstrated characteristic
sign changes associated with particle-hole symmetry,
as well as the emergence of a Fano-like profile at higher energies
for strong spin polarization and large values of the correlated hopping.
Similarly, the spin thermopower exhibited a robust dip-peak structure
whose amplitude increased with spin polarization,
but became suppressed and shifted toward higher energies
under the influence of correlated hopping.
In addition, we have also found a lack of symmetry upon changing 
$x\rightarrow 2-x$, which is associated with the presence of exchange field.
Altogether, the presented results offer further valuable insights
into the interplay between spin-dependent effects and electronic correlations,
associated with correlated hopping and Kondo effect in particular,
which could be important for guiding future experimental studies
and for potential applications in thermoelectric quantum devices.

\medskip
\textbf{Acknowledgments} \par 
This work was supported by the National Science Centre in Poland
\\
through the project No. 2022/45/B/ST3/02826.

\medskip

%

\textbf{References}\\

\printbibliography

\end{document}